\documentclass[3p,times,procedia]{elsarticle}
\usepackage{ecrc}
\usepackage[bookmarks=false]{hyperref}
    \hypersetup{colorlinks,
      linkcolor=blue,
      citecolor=blue,
      urlcolor=blue}
\volume{00}

\firstpage{1}

\journalname{Procedia Computer Science}

\runauth{M. Petrescu and T. Mihoc}

\jid{procs}

\usepackage{amssymb}

\usepackage{graphicx}
\usepackage{color}
\usepackage{array}
\usepackage{microtype}
\usepackage{booktabs}
\usepackage{caption}
\usepackage{subcaption}
\usepackage{url}
\usepackage[dvipsnames]{xcolor}

\usepackage[figuresright]{rotating}
\begin{document}
\begin{frontmatter}



\dochead{28th International Conference on Knowledge-Based and Intelligent Information \& Engineering Systems (KES 2024)}%
%% e.g. \dochead{17th International Conference on Dynamical Processes in Excited States of Solids}

\title{Massive Online Course on Entrepreneurship. Case Study}


\author[a]{Manuela Petrescu} 
\author[a]{Tudor Dan Mihoc}{cor1}

\address[a]{Department of Computer Science, Babe\c s Bolyai University, Cluj-Napoca, Romania}

\begin{abstract}
Entrepreneurship is a key component of society, and universities and major political structures have tried to support its development in recent years. The present study aims to check the perception of students (based on gender) about entrepreneurial intentions after participating in a course that had a large number of undergraduate students. There were 970 students enrolled from different faculties with various specializations. We conducted a gender-based survey on the unconventional entrepreneurial fundamentals course, where each course was delivered by a different speaker. We also compared the responses provided by computer science students with the overall responses to find differences in their perceptions related to the feasibility of teaching entrepreneurship online, determining the entrepreneurial intention of the students taking this course, and analyzing the perceptions related to the business environment and the ease of starting a business. We found that students, regardless of gender or field of study, prefer interactive online presentations based on the manner in which lectures on this subject were conducted.
\end{abstract}

\cortext[cor1]{Corresponding author. Tel.: +40-740330982.}
\end{frontmatter}

%\correspondingauthor[*]{Corresponding author. Tel.: +0-000-000-0000 ; fax: +0-000-000-0000.}
\email{tudor.mihoc@ubbcluj.ro}

% \linenumbers


%\enlargethispage{-7mm}
\section{Introduction}

Entrepreneurship is a key component of society, and universities and major political structures have tried to support its development in recent years. Scholars and practitioners are still looking for the best methods to teach entrepreneurship, as an entrepreneur is expected to have different skills: leadership, attention to detail, the ability to develop a business plan, pitch a project idea, find required resources (financial and human resources), and cope with specific risk, just to name a few of them.  
The European Union in \textit{Enterprise 2020 Action Plan} %\footnote{\url{https://ec.europa.eu/growth/smes/supporting-entrepreneurship/entrepreneurship-education/commissions-actions-entrepreneurship-education_en}}
developed programs to exchange best practices and promote entrepreneurial education. There are even programs created for young children to learn about the entrepreneurial mindset \cite{pacco2011teaching} and at university levels, more and more courses teach and promote entrepreneurial education \cite{kuratko2005emergence,Wagner2019,Moses2019} in normal circumstances or pandemics \cite{educsci1}. The best universities have created organizations for supporting women in business, for developing their entrepreneurial skills; just to name a few: Harvard Undergraduate Women in Entrepreneurship Organization (HUWE) \cite{harvardB}, Yale Women in Business Organization (YWIB) \cite{yaleB}, or Cambridge University Women in Business Organization \cite{camwib}.

The present study proposes to check the perception of students (based on gender) about entrepreneurial intentions after participating in a course that had a large number of undergraduate students from different faculties with various specializations. 
The contributions of this study are threefold. 
First, we proposed teaching entrepreneurship in an original format, in which the university decides the curricula, and the lectures are taught by people who have domain expertise in that specific area, with a focus on course consistency and accessibility. Moreover, a computer science teacher from the department of Computer Science was involved in each lecture as a moderator to facilitate the interaction between the students and the domain expert. Second, we asked for anonymous feedback from students after each course to verify their level of understanding, and at the end of the course, we asked not only their opinion but also their gender.
Finally, the responses were analyzed and discussed. 

This study was carried out considering the following objectives: 
\begin{itemize}
    \item [] 1. To check the feasibility of teaching entrepreneurship online to a large number of students enrolled in different faculties by analyzing the student's challenges;
    \item [] 2. To determine students' entrepreneurial intention;
    \item [] 3. Analyze perceptions related to the business environment and the ease of starting a business for men vs. women.
\end{itemize}
For all of the proposed objectives, we used a gender-based approach and a comparison between computer science students and students from other faculties enrolled in the course. 

These objectives can be achieved through three research questions: 
\begin{itemize}
    \item []\textit{\textbf{RQ1:} What motivated the students to sign up for the entrepreneurship course? }
    \item []\textit{\textbf{RQ2:} Do students think the business environment is more favorable for women or men, and why?}
    \item []\textit{\textbf{RQ3:} What were the Course's Challenges?}
\end{itemize}

The rest of the paper is organized in the following manner: we discuss recent approaches related to gender and entrepreneurship in the Literature Review Section, followed by the Methodology Section, where we describe the methodology used for this case study. In the Data Collection Section, we describe how we collected the data. Next, we investigate the answers to the surveys and research questions in Results and Discussions. In the Threats to Validity Section, we address the threats to validity and conclude in the Conclusion Section.

\vspace{-5pt}
\section{Literature Review}
\label{sec:literature_review}
Most industrialized nations encourage diversity and inclusion because it is beneficial to society to integrate as many people as possible. In science, as in business, women can be innovative, hard-working, and can bring value to the domain to which they contribute \cite{Leisyte2021,Ellemers2022}.

Universities are exploring the best methods for teaching entrepreneurship. There is an interest in teaching entrepreneurial skills to women to minimize the gap between different genders in the business domain. Online experiences of teaching entrepreneurial skills at Lancaster University during the COVID pandemic are described in \cite{educsci1}, and other articles discuss gender characteristics in entrepreneurial education in Eastern and Northern Europe \cite{Tatyana2019} or the gender gap in tech entrepreneurship \cite{Wilson22}. Women are underrepresented in Science, Technology, Engineering, and Mathematics (STEM), and also in entrepreneurship, so experiments involving women and entrepreneurship learning are performed in terms of entrepreneurial self-efficacy and entrepreneurial intent \cite{Elliott20}. In "Gender and Entrepreneurship Research: A Review of Methodological Approaches" the authors argue about developing new teaching methods to match the trends to suit women better \cite{Henry2015}. Ioannis Sitaridis et al. considered that entrepreneurial intentions in the information technology field are understudied, so they conducted a study related to "the impact of gender typed personality and social norms in conjunction with the role of entrepreneurial education" and concluded that personality traits associated with masculinity are strongly related and positively correlated with entrepreneurial intentions \cite{Sitaridis2018}.  Haus et al. obtained a similar conclusion: men's intentions are on average higher than women's intentions. They conducted a review study of 30 research articles related to the effect of gender on entrepreneurial intentions and concluded that the difference between genders is, in fact, "a consequence of differences in turning intentions into implementation” \cite{haus2013}. Subjective well-being and performance seem to be correlated with gender and different types of activity: women achieve "greater subjective well-being and levels of performance when they are high in creativity" \cite{HMIELESKI2019709} and men are better at teamwork. However, both positive gender incongruities are beneficial to all parties involved (men and women) \cite{HMIELESKI2019709}. 

\vspace{-5pt}
\section{Methodology}
%The methodology adopted in this study is presented in this section. 

\subsection{Participants}

In the entrepreneurial course, 970 students from 15 faculties were enrolled, with different backgrounds, skills, and interests. Due to this diversity, the lectures had to be adapted to be understood by everyone. The proposed examples had to be clear and generic, as close as possible to everyday life and their experiences. Usually, students prefer familiar examples rather than \textit{"waw examples" (the most, the worst...)} \cite{MP}. As digitization is a key factor in a company's success, the examples were mainly from this domain, but they were presented in such a manner that even students without any computer science knowledge can understand and assimilate the concepts and the explained software (the lectures contained familiar examples, pictures, and they used a less technical language).

To better understand the student's mindset, we checked what faculties they come from. 
Using university public data \cite{UBBstat2024}, we calculated the percentages of students enrolled in each faculty using the formula: the number of students enrolled in the course divided by the number of students in the faculty.
The statistic is presented in Table \ref{tab:enrolled_students}. 

We noticed an interesting fact: the highest number of enrolled students studied at the Faculty of Physical Education and Sport. Later, we found the reason: this faculty was the only one that automatically enrolled students from a line of study; for all the other faculties in the university, enrollment was optional.

As the course was optional, we concluded that the enrollment rates were showing some interest; however, a further inquiry will be necessary to determine whether this interest is based on field dynamics or on the ease of opening a start-up in the business domain. The second largest group in terms of number of students was formed by students from the Faculty of Mathematics and Computer Science, and it may be correlated with the large number of start-ups in the IT domain (a modern trend). Finally, we noticed that some faculties were represented only by a few students (4 students from the Faculty of Law, one from the Faculty of Letters, and 7 from the Faculty of Sociology and Social Work). 

We observe that students associated with faculties associated with STEM fields, as well as theology faculties, show greater interest in this field. Furthermore, the percentage of students interested in theology is more than twice that of STEM (10\% vs 4.9\%). The level of interest was not the same among students from other humanities faculties. We reiterate that we did not take the sports faculty into consideration for the previously mentioned reasons.

\begin{table}
\centering
\caption{Number of students and percentage of students that choose to enroll to the course per faculty. The 15\% for the Sport Faculty is not significant due to the automatic enrollment of some students.}
\begin{tabular}{ | l | c |c | }
\hline
\textbf{Faculty} &\textbf{ No. of students} & \textbf{Percentage}\\
\hline
	Mathematics and Computer Science & 146.0 & 4.50 \% \\ \hline
	Physics & 13.0 & 4.90 \%\\ \hline
	Chemistry & 31.0 & 2.80 \% \\ \hline
	Biology and Geology & 18.0 & 1.30 \% \\ \hline
	Law & 4.0 & 0.10 \%\\ \hline
	Letters & 1.0 & 0.03 \% \\ \hline
	History and Philosophy & 13.0 & 0.80 \% \\ \hline
	Sociology & 7.0 & 0.30 \% \\ \hline
	Psychology & 29.0 & 0.40 \% \\ \hline
	Business & 90.0 & 5.00 \% \\ \hline
	Political Science, Administration  and Communication &86.0 & 2.10 \%\\ \hline
  	Sport & 251.0 & 15.6 \% \\ \hline
	Greek Catholic Theology & 26.0 & 6.80 \% \\ \hline
	Romano Catholic Theology & 30.0 & 10.00 \% \\ \hline
	Environment Engineering  & 1.0 & 0.16 \% \\ \hline
\end{tabular}

    \label{tab:enrolled_students}
\end{table}

\subsection{Experimental Setup}

The large number of students presented some organizational challenges that supported the decision to use an online format for the course. 
Their different skills and specializations forced educators to adapt the content to be understood by a large and diverse group of students. At the end of each course, there was a live Q\&A session moderated by the computer science teacher. Each lecture was recorded and uploaded to a dedicated platform to address the possibility of missing a course. 
The lecture presentations were also available (after the course had been delivered) and could be accessed via the course website, making information more accessible to all interested students.

The contingency approach to recording and making it available later was appreciated by the students. In one of the feedback responses we received the following comment: \textit{"I found important information, I need to watch some of the lectures again"}.

\subsection{Lecture setup}
\label{sec:lecture}

In pursuit of improvement, to lower the dropout rates, and to involve the students, we proposed that every interaction had to be online. This includes the topic presentations, the Q\&A sessions, the grading, and all the announcements. 

Each lecture had an important digitization component: from marketing (online campaigns and social media marketing and ads) to the legislation aspects (how to find details on the platforms provided by the state institutions and how to check the latest financial information). We discussed aspects related to learning different software tools: how to use them to optimize processes or how to create a business plan. After each lecture, we asked for feedback and the students had to take a test. Therefore, we could evaluate student participation, their understanding of the topics taught, and their feelings and perceptions related to the course in general and each topic in particular. 
Due to some constraints (large number of students and timetable restrictions), we proposed to use the online approach. All the interactions, starting with the announcement, continuing with lectures and Q\&A sessions, emails, messaging, and grading were online.
The schedule presented a challenge because we had to accommodate different constraints, so we chose a time slot that was convenient for the vast majority of students.

The gender-based questions were only asked in the final feedback form.

\subsection{Course structure}
\textbf{}{Fundamentals of Entrepreneurship} was an optional course organized at the university level.
The course contained 12 lectures of two hours each; in each lecture, half an hour was allocated for Q\&A sessions. 
We chose multiple speakers with entrepreneurial experience, helped by a moderator from the Computer Science department. In the unlikely event of scheduling conflicts or other unexpected events, a university backup teacher was prepared to deliver the course.
The speakers manage or have a management role in IT companies, IT hubs, or IT innovation clusters; therefore, the examples provided and explained in the course were related to the computer science domain. 

We establish the curriculum in accordance with the syllabuses for the same course from other universities by a commission formed by teachers from the computer science department.
The resulting curriculum is mentioned in Table~\ref{tab:curricula}.

\begin{table}
\centering
  \caption{\textit{Fundamentals of Entrepreneurship} curriculum}
  \label{tab:curricula}
  {\small{
  \begin{tabular}{|lll|}

\hline
Lecture 1. & & Introduction to Entrepreneurship and Innovation. \\
Lecture 2. & & From problem to idea. Validation of the solution. Unique selling proposition. \\
Lecture 3. & & Digital Transformation and Innovative Thinking. \\
Lecture 4. & & Who are my Potential Clients and what are their expectations? \\
Lecture 5. & & Who are my competitors and how can I gain a Competitive Advantage? \\
Lecture 6. & & Teams and leadership. How to build a team? Team roles. \\
Lecture 7. & & Business plan. \\ 
Lecture 8. & & Social responsibility. \\
Lecture 9. & &  Minimum Viable Product (MVP). Launching a product. \\
Lecture 10. & &  Public speaking. Preparation and presentation of a pitch. \\
Lecture 11. & & Entrepreneurial ecosystem and growth opportunities. \\
Lecture 12.& &  Funding opportunities. \\
\hline
\end{tabular}
}}
\end{table}

\section{Data Collection}
\label{sec:data_collection}

After each lecture, we asked for feedback and asked them to rate the material and, if they wanted, to provide a comment to support their rating. This is the process in which we gather the data. At the end of the course, a more complex survey with open-ended and accountable questions was given. We chose to use open-ended questions because they provide a deeper insight into the opinions and perceptions of the students, while accountable questions give us a simpler way to analyze the students' characteristics and categorize them accordingly. 

The survey questions were translated into two languages: Romanian and English, allowing each student to select their preferred language. Responses written in Romanian were automatically translated to English using translation tools. The authors then verified the translated responses, checking for language flavors and other inconsistencies.

We used online tools for data collection and did not ask for data that included personal information such as names or emails. Also, we did not request that the faculty appear, because the small number of students from some faculties would have given them the impression that we could have identified them. We instead asked for the gender.

The methods used are quantitative methods and more specific questionnaire surveys as defined in community standards \cite{ACM}, the methods were previously used in \cite{tichy,redmond2013}.

The questionnaire was open for two weeks at the end of the course, and 123 students participated and provided answers. The survey questions relevant to this study are shown in Table \ref{tab:course_feedback}.

\begin{table}[h]
\centering
  \caption{Queries from the course completion feedback survey}
  \label{tab:course_feedback}
  {\small{
  \begin{tabular}{|lll|}
\hline
(1) & & Specify your gender \\
(2) & & Specify your faculty \\
(3) & & Why did you decide on this particular course? \\
(4) & & What did you hope to learn from this course? \\
(5) & & What do you consider the most important thing / things you have learned in this course? \\
(6) & & Do you think there is a difference between the way women can do business?\\
(7) & & Do you think the business environment is more favorable for women or men?  Why? \\
 \hline
\end{tabular}
}}
\end{table}

There were 45.85\% women enrolled in the course and 54.15\% men. In our survey, the percentage of women who responded was 45.28\%, and the percentage of men was 54.71\%. When comparing the percentages of the responses, we noticed that the numbers were close to each other; thus, we conclude that the answers provided in the study were relevant for all the students enrolled. 
 
\section{Results and Discussions}
The following findings for each research question are obtained after collecting data and thematic analysis.

\subsection{RQ1: What motivated the students to sign up for the entrepreneurship
course?}
Being an optional course (with the exception of the Sports Faculty), we were interested in the main reasons why the students enrolled and whether there was a gender-based difference in their approaches. For this, we asked students questions (3) and (4) from Table \ref{tab:course_feedback}. 

\begin{figure}[h]
    \centering
\begin{subfigure}[b]{0.45\linewidth}
  \includegraphics[width=\linewidth]{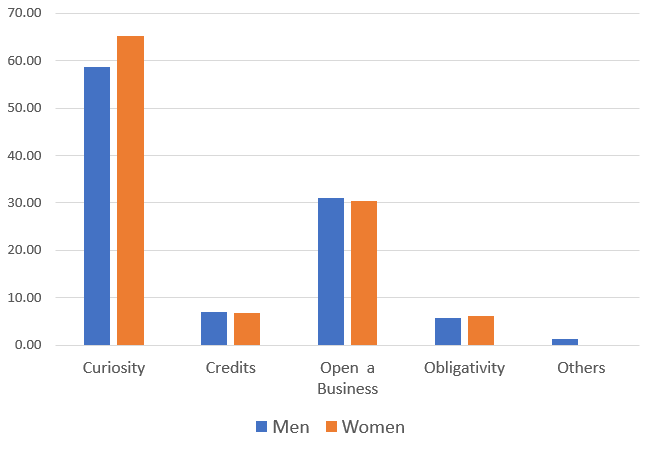}
  \caption{Total of enrolled students}
  \label{fig:motivation}
\end{subfigure}
\hfill
\begin{subfigure}[b]{0.45\linewidth}
  \includegraphics[width=\linewidth]{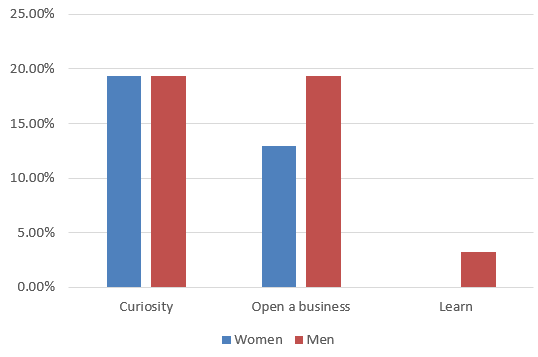}
  \caption{Computer Science students}
  \label{fig:motivationCS}
\end{subfigure}
    \caption{Students' reasons for enrolling in the \textit{Fundamentals of Entrepreneurship} course. Gender-Based}
    \label{fig:enter-label}
\end{figure}

There were no significant differences related to gender when answering these questions (the responses that did not specify their gender were not included), as can be seen in Figure 
\ref{fig:motivation}, and the reasons why women and men enrolled in this course are relatively similar. The same observation applies to computer science students, see Figure \ref{fig:motivationCS}. There was a small number for other reasons: for example \textit{"I enrolled by chance and I wanted to finish the course"}.
Most of the students who answered the first question with the option \textit{"I want to start a business"}, reinforced their intention when answering the second question: \textit{"I hoped to learn business models and development strategies"}, \textit{"Methods to finance my company"}, \textit{"business ideas"}, \textit{"entrepreneurial mindset"}.

\subsubsection{Relation between Curiosity and Motivation}
\begin{figure}[h]
  \centering
  \includegraphics[width=0.56\linewidth]{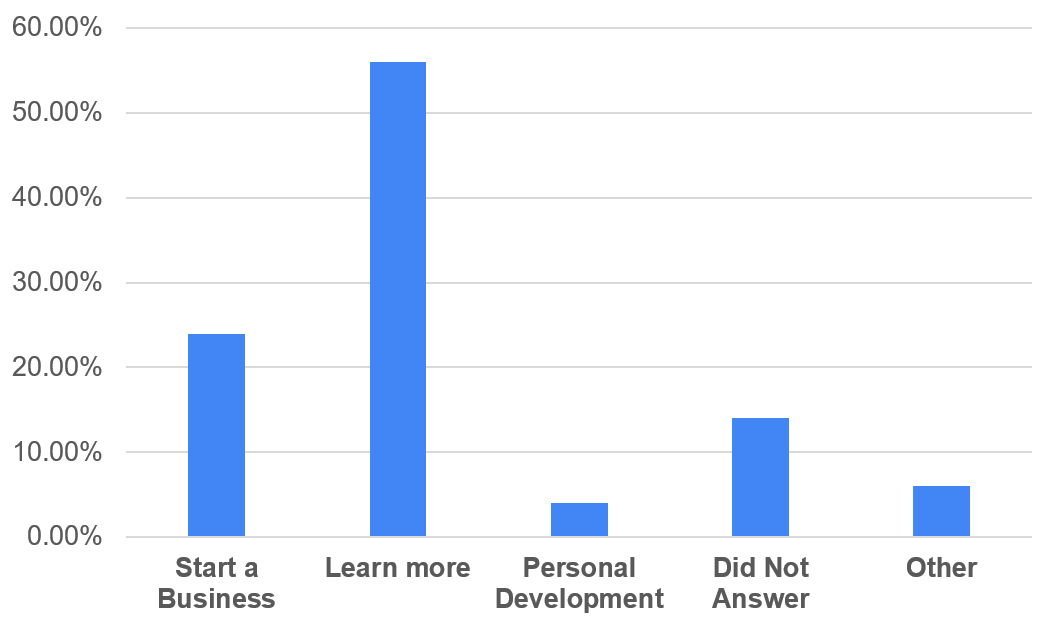}
  \caption{Students expectation for those who identified \textit{Curiosity} as the main reason to enroll in the course.}
  \label{fig:curiosityRelation}
\end{figure}

Most of the students answered \textit{ Curiosity}, followed by \textit{"I want to start a business"}. However, analyzing the responses, we noticed that these two items are connected, as most of the \textit{curious} students replied to the second question with statements such as \textit{"How to start a business"}, \textit{"To find information that could help me in the future"}, \textit{"How to manage a business"}. The students who specified \textit{curiosity} as the main reason and their expectations are presented in Figure \ref{fig:curiosityRelation}.  

In conclusion, the student's interest in the course was mainly due to the intention to open a business, and second, to their curiosity and desire to learn.

\subsection{RQ2: Do students think that the business environment is more favorable for women or men, and why?}
Even if, from a legal point of view, there is no gender-based difference, the student's responses reflect a different point of view. We analyzed the responses, and the majority of the students appreciated that men and women have equal chances and opportunities and that the business environment is as favorable for men as it is for women. However, other responses appreciated that men have more opportunities and are perceived as more suitable for managerial / business roles. We also compared computer science students' perceptions versus overall perception related to this question. 

\begin{figure}[h!]
\centering
\hspace{40pt}
   \begin{subfigure}[b]{0.45\textwidth}
          \includegraphics[width=0.8\textwidth]{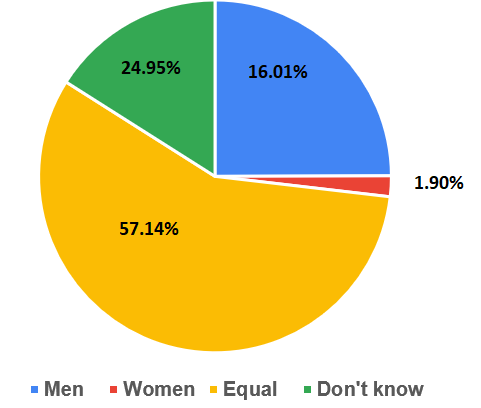}
         \caption{Total of enrolled students }
         \label{fig:fig_skills}
   \end{subfigure}
%\hfill
   \begin{subfigure}[b]{0.45\textwidth}
         \includegraphics[width=0.75\textwidth]{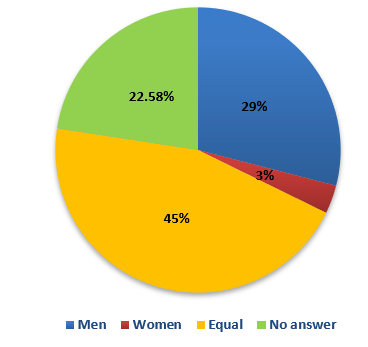}
         \caption{Computer Science Students}
         \label{fig:fig_skillsCS}
   \end{subfigure}
   \caption{The percent of students that perceive gender as an advantage in the business environment.}
\end{figure}

The percentage of students in computer science who appreciated that the business environment is more favorable to men is higher than the general percentage of 29\% versus 16\%, and the percentage of students who appreciated that men and women have equal chances is smaller: 45\% versus 57.14\% as can be seen in Figure \ref{fig:fig_skillsCS}. 

From the responses received from the 123 students, synthesized in Figure \ref{fig:fig_skills}, we can draw the following observation about the power-related attributes. 

\begin{figure}[h]
  \centering
  \includegraphics[width=0.56\linewidth]{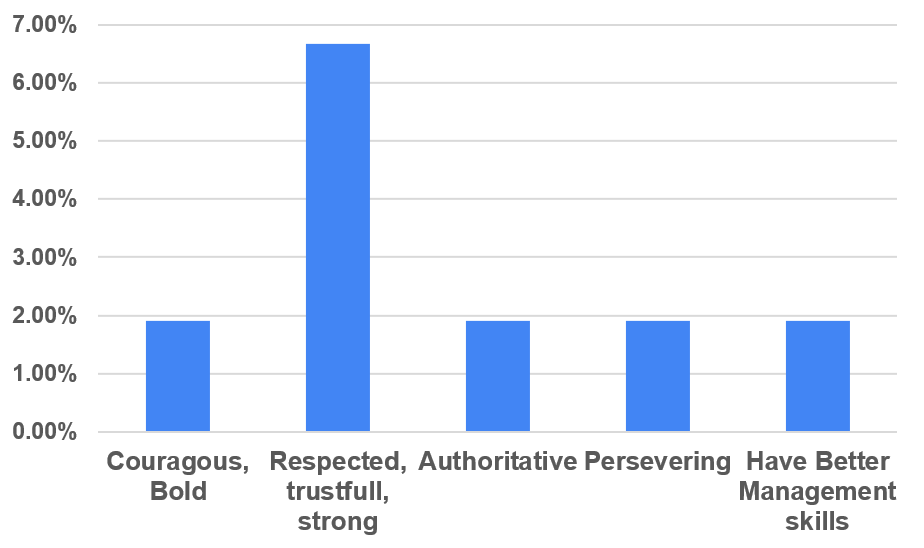}
  \caption{Men's Characteristics in students' opinions}
  \label{fig:men_skills}
\end{figure}

A set of characteristics was mentioned in the open responses received: courage, boldness, general thinking, management skills, strength, respect, trustfulness, authority, and perseverance. These characteristics were mostly assigned to men. In their collective perception, 24.95\% of the students state that men are more suitable to start and run a business. Sometimes they offer motivation; in other cases, their answer to \textbf{RQ2} questions is straightforward without additional explanation: \textit{For men, Men} or \textit{Men, for obvious reasons}. 
In Figure \ref{fig:men_skills}, the percentage represents the student's opinion related to men. These students consider men to be more courageous, bold, or have better management skills. Some answers are based on personal observations \textit{"Unfortunately I notice a men's advantage due to a set of factors", "Men are more respected in this environment (business)",} and \textit{"People tend to listen to men more than women. Why? I don't know"}. Other responses were: \textit{"Men are more intelligent"} and \textit{"Men take more risks and are more courageous in unpredictable situations"}. We noticed that men were characterized only in a positive way, as women were characterized in a positive and also negative way \textit{"less capable", "unskillful", "more persevering", "more convincing"}.

Some students, 16.01\%,  decided not to answer this question or express an opinion, \textit{Don't know / no answer}. We consider them to be not interested in the topic. 

In conclusion, the perception that men have more advantages is deeply rooted; some of the students just acknowledge it, and others offer reasons based on men's capabilities and overall perception.

\vspace{-5pt}
\subsection{RQ3: What were the Course's Challenges?}
\vspace{-5pt}

The open answers provided to this question could be grouped into a few topics: \textit{there were no challenges},  \textit{time table}, \textit{course's questionnaires}, 
\textit{provided information,}
 \textit{delivery method},
 \textit{did not answer}.
 
\begin{figure}

\centering
\begin{subfigure}[b]{0.48\linewidth}
  \includegraphics[width=\linewidth]{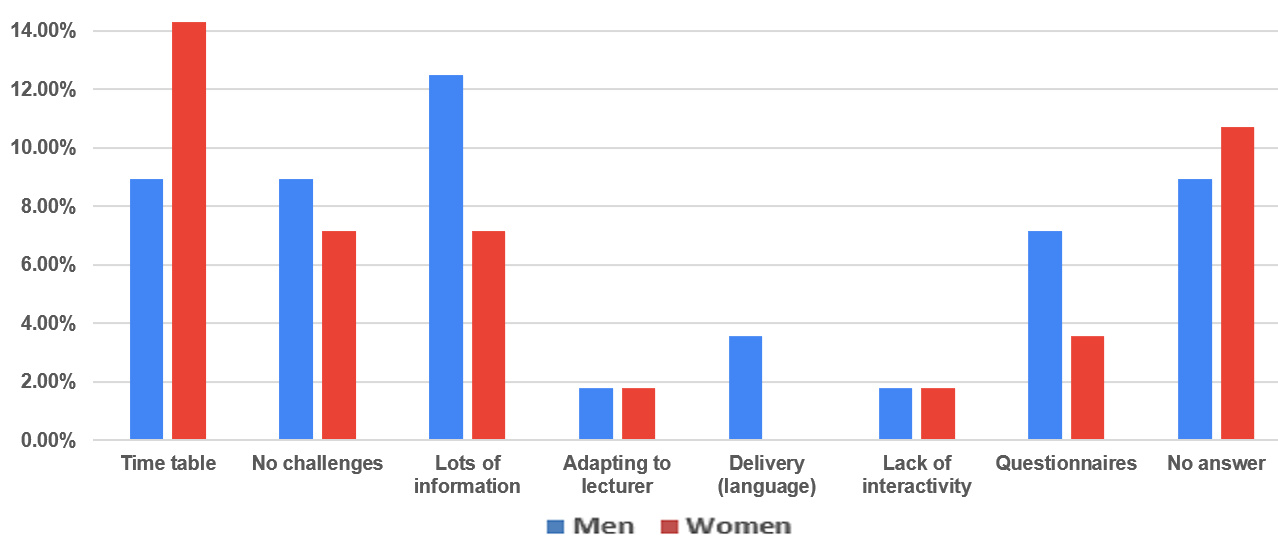}
  \caption{Total enrolled students}
  \label{fig:challenges}
\end{subfigure}
\hfill
\begin{subfigure}[b]{0.48\linewidth}
  \includegraphics[width=\linewidth]{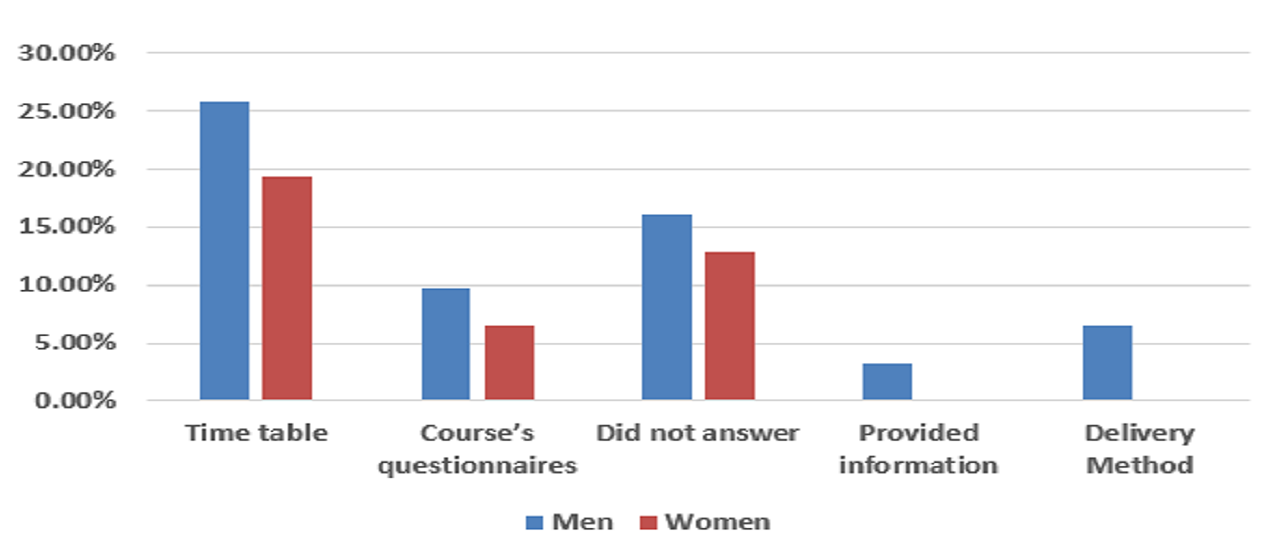}
  \caption{Computer Science students}
  \label{fig:challengesCS}
\end{subfigure}
 \caption{Challenges for the enrolled students by Gender}
\end{figure}

The gender-based challenges for all the enrolled students are presented in Figure \ref{fig:challenges} and we can observe that, in part, they are similar to the challenges mentioned by the computer science students (Figure \ref{fig:challengesCS}). 

A large group of students appreciated that they did not encounter any difficulties or challenges, others did not respond to this question. More than 50\% of the students are in these two groups.

For us, one of the major challenges of this course was the timetable. With a large number of students enrolled, it was difficult to find a time slot for everyone to be available, even if the course had an online format. 
The hour during which the lecture was held was also a major issue raised by some of the students, even if we specifically chose to deliver this course late in the evening. However, taking into account these aspects, as well as the number of students who did not face any challenges or respond, we concluded that most students did not consider timing as a problem.

Students mentioned that they encountered difficulties related to each evaluation of the lecture. Part of the challenge was due to the fact that they could not participate in the course and part of it was due to the different teaching methods of each speaker. Regarding this aspect, we informed all speakers that the course would be generic and that they would have to adapt to a diverse group of students. Other challenges were mentioned by a smaller number of people: adapting to each lecturer, to the delivery method, \textit{"too many teaching styles", "language"}. Technical issues were also mentioned: \textit{"I could not hear well"}.

Some students appreciated that the course contained \textit{"too much information", "many new terms", "unknown words", "the financial part"}. It was interesting to note that the percentage of men who mentioned this challenge was higher than the percentage of women.

\vspace{5pt}
\textbf{Discussions:}
Upon examining these findings, we note the following broad features: 
\vspace{5pt}

Since numerous speakers were used, this led to a number of unfavorable outcomes and challenges related to the ongoing adaptation to different teaching philosophies. The large, massive online course format was well-received by the students, in part because of its accessibility. Given that some of the material covered in the course can be challenging, the interactive Q\&A section at the end of the course was important to the students. This is a feature that is absent from the recorded courses. In conclusion, regarding the first objective of our study: It would be feasible to hold the course online in a traditional interactive format, and it is preferable to a recorded one or with many invited speakers.

For the second objective, we see that students in theology departments, as well as those from STEM departments, are interested in entrepreneurship. Surprisingly, students in theology show the largest interest; maybe a further study is necessary to investigate the reasons. Students from different humanities studies showed lower degrees of interest.

No differences were observed between the interest of women and men in this course. They chose the course primarily out of curiosity and a desire to launch a business. In conclusion, with regard to the second objective of our study, we conclude that there are no significant differences between genders related to entrepreneurial intentions because students are driven by curiosity and the desire to start a business.

Regarding the third objective of the study, we observe that power-related attributes are associated with men by respondents of both genders.

Given the large number of participants, their diversity (located in different locations) and the fact that certain individuals state that "the general perception is that ...", we can conclude that the study reflects the view of an important segment of the population that believes that men are better at management and business. 

\vspace{-10pt}
\section{Threats to Validity}
\label{sec:ThreatsToValidity}

\vspace{-5pt}
Questionnaire studies must address threats to validity and according to existing ACM standards guidelines \cite{ACM} we decided to take into account the following: target population and sampling (participant  selection), actions taken to limit dropouts. Attention was also focused on research ethics.

Target population and participant selection: all students enrolled in the course could take part in the intermediary surveys as well as in the final one based on their own decisions. Thus, participation was neither limited nor constrained by any rules, and we addressed as many students as possible with the purpose of having relevant results. The sample group was dynamic (each intermediary survey as the last survey had different participation), so there are no threats related to a small target population or to constraints or rules that would limit the participation. As for external validity: there is a risk that the results obtained in the third objective of the study are not entirely representative for the entire population.

Dropouts: As the course had around 970 enrolled students, it was impossible to find a time slot when everyone would be available. Some students had mandatory courses in the time slot, so we faced an initial dropout even if the course was placed Thursday from 6 to 8 p.m. The selected time frame was chosen to increase the chances that students are available. Moreover, the presence was not mandatory, and we recorded and uploaded (with free student access) all the course materials (presentations and recordings) in the event someone was not able to attend; so he or she could find the relevant information. We made all necessary efforts to minimize dropouts.

Research ethics: students were informed about the reasons we collect data and their participation in feedback surveys for each course; the final questionnaire was optional, and we did not use any methods to enforce greater participation. The feedback surveys and the final questionnaire were anonymous and the answers to the questions were not mandatory. We had students who skipped some questions, we had students who provided explicit and detailed feedback or answers to specific topics (they could write as much as they wanted, as the form responses were not limited to a specific number of characters). 

\vspace{-10pt}
\section{Conclusions and Future Work}
\label{sec:conclusion}

\vspace{-8pt}
We conducted a gender-based study in a massive course on unconventional entrepreneurship fundamentals. The course was held online, there were a large number of students enrolled from different faculties, and each lecture was delivered by a different speaker. At the end of the course, we conducted a survey to find answers to the research questions. The percentage of men/women from the students who responded to the survey (123 responses) was similar to the percentage of men/women enrolled in the course. Furthermore, the students who responded to the survey were from all faculties, so we conclude that the responses are representative. We took into account a gender-based approach and also compared the responses provided by computer science students with the overall responses to find differences in their perceptions.

We came to the conclusion that students, regardless of gender or field of study, prefer interactive online presentations in a classical setting based on the manner in which lectures on this subject were conducted. 

Analyzing how the enrolled students were distributed among the faculties revealed that students studying theology and STEM fields were much more interested in entrepreneurship than students studying the humanities. Further research on the motivations for why students in theology make an oddity out of the other ones in the humanities would be interesting. 

We learned about the difficulties the students encountered: time constraints, presentations, difficulties adapting to multiple speakers, or evaluations. 
There were no significant differences in the answers provided by the computer science students compared to the rest of the answers related to the course challenges.

We found that most students do not perceive that the business environment benefits men (57.14\% from all answers compared to 45\% from the computer science students' answers). However, 24.95\% of all students and 29.03\% of computer science students stated that men are perceived to be advantaged, perhaps because some of society has the perception that men are better at doing business. The open questions provided a more in-depth analysis; some students considered that men and women have the same business opportunities. 

In the future, we want to see if business perception changes after taking this course (gender-based), and how they understand and perceive the digitization examples provided in the course. It would also be interesting to find out how they would implement digitization in their field of expertise. Future research should examine whether showcasing successful female entrepreneurs can change public opinion regarding the role women play in entrepreneurship.
Some practical suggestions for future improvements of this particular course include a smaller number of teachers who follow a similar teaching style, as well as arranging groups taking into account the students' various levels of knowledge. This can be determined through a survey administered at the beginning of the course about the concepts that will be presented. The material can thus be modified on the basis of the results.

\vspace{-10pt}

\bibliographystyle{splncs04}
\bibliography{kes24.bib}

\begin{thebibliography}{10}
\providecommand{\url}[1]{\texttt{#1}}
\providecommand{\urlprefix}{URL }
\providecommand{\doi}[1]{https://doi.org/#1}

\bibitem{UBBstat2024}
Statistics data {U}{B}{B} (accessed in january 2024), \url{"https://www.ubbcluj.ro/ro/infoubb/documente_publice/statistici"}

\bibitem{camwib}
Cambridge university women in business organisation (accessed September 2022), \url{"https://www.camwib.com/?fbclid=IwAR0kMe5kJg_vzpFwq77Sz0Va7ruqp9ql_0TtnxMJ70gEdtgduUOlqepTUzQ"}

\bibitem{harvardB}
Harvard undergraduate women entrepreneurship organization (accessed September 2022), \url{"https://thehub.college.harvard.edu/organization/huwe"}

\bibitem{yaleB}
Yale women in business (ywib) organization (accessed September 2022), \url{"https://campuspress.yale.edu/business/"}

\bibitem{Ellemers2022}
Ellemers, N., de~Gilder, D.: Diversity and inclusion. In: The Moral Organization: Key Issues, Analyses, and Solutions, pp. 161--200. Springer (2022)

\bibitem{Elliott20}
Elliott, C., Mavriplis, C., Anis, H.: An entrepreneurship education and peer mentoring program for women in stem: mentors’ experiences and perceptions of entrepreneurial self-efficacy and intent. International Entrepreneurship and Management Journal  \textbf{16},  43–67 (2020). \doi{10.1007/s11365-019-00624-2}

\bibitem{haus2013}
Haus, I., Steinmetz, H., Isidor, R., Kabst, R.: Gender effects on entrepreneurial intention: a meta-analytical structural equation model. International Journal of Gender and Entrepreneurship  (2013)

\bibitem{Henry2015}
Henry, C., Foss, L., Ahl, H.: Gender and entrepreneurship research: A review of methodological approaches. International Small Business Journal  \textbf{34} (01 2015). \doi{10.1177/0266242614549779}

\bibitem{HMIELESKI2019709}
Hmieleski, K.M., Sheppard, L.D.: The yin and yang of entrepreneurship: Gender differences in the importance of communal and agentic characteristics for entrepreneurs' subjective well-being and performance. Journal of Business Venturing  \textbf{34}(4),  709--730 (2019). \doi{10.1016/j.jbusvent.2018.06.006}

\bibitem{kuratko2005emergence}
Kuratko, D.F.: The emergence of entrepreneurship education: Development, trends, and challenges. Entrepreneurship theory and practice  \textbf{29}(5),  577--597 (2005)

\bibitem{educsci1}
Lambert, C.G., Rennie, A.E.W.: Experiences from covid-19 and emergency remote teaching for entrepreneurship education in engineering programmes. Education Sciences  \textbf{11}(6) (2021). \doi{10.3390/educsci11060282}

\bibitem{Leisyte2021}
Leisyte, L., Deem, R., Tzanakou, C.: Inclusive universities in a globalized world. Social Inclusion  \textbf{9}, ~1--5 (07 2021). \doi{10.17645/si.v9i3.4632}

\bibitem{Tatyana2019}
Monastyrskaya, T.I., Shved, V.V., Chudinov, S.I., Medvedeva, N.P., Mazurkevich, O.P.: Teaching entrepreneurship and gender-based assessment of entrepreneurial competences. International Journal of Recent Technology and Engineering (IJRTE)  \textbf{8}(3C) (2019). \doi{10.35940/ijrte.C1017.1183C19}

\bibitem{Moses2019}
Olokundun, M.A., Moses, C.L., Iyiola, O., Ogunaike, O., Ibidunni, A.S., Kehinde, O., Motilewa, D.: Experiential pedagogy and entrepreneurial intention: A focus on university entrepreneurship programmes (2019)

\bibitem{pacco2011teaching}
Pa{\c{c}}o, A.d., Palinhas, M.J.: Teaching entrepreneurship to children: a case study. Journal of Vocational Education \& Training  \textbf{63}(4),  593--608 (2011)

\bibitem{MP}
Petrescu, M., Sterca, A.: Agile methodology in online learning and how it can improve communication. a case study. In Proceedings of the 17th International Conference on Software Technologies pp. 542--549 (2022)

\bibitem{ACM}
{Ralph, Paul (ed.)}: {ACM Sigsoft Empirical Standards for Software Engineering Research}, version 0.2.0 (2021), \url{https://github.com/acmsigsoft/EmpiricalStandards}

\bibitem{redmond2013}
Redmond, K., Evans, S., Sahami, M.: A large-scale quantitative study of women in computer science at stanford university. In: Proceeding of the 44th ACM technical symposium on Computer science education. pp. 439--444 (2013)

\bibitem{Sitaridis2018}
Sitaridis, I., Kitsios, F.: Entrepreneurial intentions in the field of it: The role of gender typed personality and entrepreneurship education. In: 2018 IEEE Global Engineering Education Conference (EDUCON). pp. 1854--1859 (2018). \doi{10.1109/EDUCON.2018.8363460}

\bibitem{tichy}
Tichy, W.F., Lukowicz, P., Prechelt, L., Heinz, E.A.: Experimental evaluation in computer science: A quantitative study. Journal of Systems and Software  \textbf{28}(1),  9--18 (1995)

\bibitem{Wagner2019}
Wagner~M., Schaltegger~S., H.E.e.a.: University-linked programmes for sustainable entrepreneurship and regional development: how and with what impact? Small Business Economics p. 1141–1158 (2021). \doi{/10.1007/s11187-019-00280-4}

\bibitem{Wilson22}
Wilson, A.W., Patón-Romero, J.D.: Gender equality in tech entrepreneurship: A systematic mapping study. In: 2022 IEEE/ACM 3rd International Workshop on Gender Equality, Diversity and Inclusion in Software Engineering (GEICSE). pp. 51--58 (2022)

\end{thebibliography}

\end{document}